\documentstyle[12pt]{article}
\let\sref=\ref
\def\ref#1{(\sref{#1})}
\def\Ten#1{\gdef\u{\vphantom{#1}}#1}
\def\Det{\mathop{\rm Det}}

\begin{document}

\begin{flushright}   FIAN/TD/95-12 \\
                hep-ph/9506132\\
                June 1995 \vspace{3ex}
\end{flushright}

\vbox to 1in{\vfill}

\hbox to\hsize{\hss\Large\bf On the Equivalence of Dual Theories\hss}

\vskip .5in

\hbox to\hsize{\hss\large A.Subbotin, I.V.Tyutin\footnote{E-mail
address: tyutin@lpi.ac.ru}\hss}

\begin{center}
\it P.N.Lebedev Physics Institute, \\
\it Leninsky Prospect 53, 117924, Moscow, Russia.
\end{center}

\vskip 1in

\hbox to\hsize{\hss\bf ABSTRACT\hss}

\vskip.5in

\leftskip 30pt
\rightskip 30pt

\noindent
We discuss the equivalence of two dual scalar field theories in 2 dimensions.
The models are derived though the elimination of different fields in
the same Freedman--Townsend model. It is  shown that tree $S$-matrices
of these models do not coincide. The 2-loop counterterms are calculated.
It turns out that while one of these models is single-charged, the other
theory is multi-charged. Thus the dual models considered are
non-equivalent on classical and quantum levels. It indicates the possibility
of the anomaly leading to non-equivalence of dual models.

\leftskip 0pt
\rightskip 0pt

\newpage

\section{Introduction}

The duality transformations are nowadays widely used in a field
theory, providing the description of physical systems on
alternative groundings. The first example of this approach is
likely to be given by the Kramers-Wannier duality~\cite{K}, relating
the low-temperature properties of lattice models with the high-temperature
ones. (See~\cite{R} for a review on the applications of duality
transformation in superstring theory, and references therein). The
original model and its dual are commonly assumed to be physically equivalent.
However, in quantum theory the transformations like changes of
variables may induce anomalies, with axial and conformal anomalies
being their typical examples. Thus, the question on equivalence of
dual theories deserves a more thorough discussion.

The present work is aimed to join this discussion, considering
two dimensional Freedman--Townsend model~\cite{T}. In the general
opinion, this model is equivalent to the model of principal chiral field
$\varphi^a$~\cite{T,Cu,Sl}. The latter one arises after elimination
of vector fields $A_\mu^a$ using the equations of motion due to
variations of the action by a scalar field $B^a$ (which an antisymmetric
tensor in 2~dimensions is reduced to). On the other hand, when the
equations of motions due to $A_\mu^a$ field variations are used, one
arrives at the theory written in terms of the field $B^a$. The two models:
the model of the principal chiral field $\varphi^a$, and that of the
field $B^a$, are related by duality transformation~\cite{FZ,F,A,Cu}.
Can one treat them as equivalent? We show in this work, that in
perturbation theory one {\sl can not.\/} To compare the models, the
Born scattering amplitudes $2\to2$, and the 2-loop counterterms are
calculated. Though even the Born amplitudes turn out to be different,
the arguments based on calculations of $S$-matrix elements could be
considered dubious, since the massless particles are involved. The
comparison of counterterms provide more powerful ones.
The geometry of the principal chiral field model constraints the
total renormalization reducing it to a multiplicative renormalization
of the coupling and a non-linear renormalization of the field~\cite{P}.
The principal chiral field model is single-charged. The $B^a$-field
model is also single-charged in one-loop approximation (it has already been
known since the work~\cite{F}), though the charge renormalizations
in these models are different. Below we demonstrate, that in two-loop
approximation the renormalization in the $B^a$-field model is {\sl not\/}
reduced to the charge and $B^a$ field renormalizations. In fact, this
model is multi-charged. The latter means that these dual models are
nonequivalent, calling for explicit checks to be made when the equivalence
of any pair of dual models is alleged.

The paper is organized as follows. In section~\sref{Formal} the
Freedman--Townsend model is described. Restricting ourselves to the
{\sl SU(2)\/} group case, we construct the action in terms of the
field $B^a$, and, after a sequence of path integral transformations,
demonstrate a formal equivalence of this model to the model of principal
chiral fields. Born scattering amplitudes are calculated in
section~\sref{2d-Born}. In section~\sref{2d-2-loops} we evaluate two-loop
counterterms. Finally, in section~\sref{3d-Born} the calculations of
Born amplitudes in two different representations of the 3-dimensional
Freedman--Townsend model are presented. In this case the amplitudes
{\sl do\/} coincide! We interpret this fact as an indication that the
possible origin of 2-dimensional anomalies lies in ill-defined
infrared behavior of the massless theory.

\section{A Formal Approach}\label{Formal}
Consider the Freedman--Townsend model~\cite{T}
\begin{equation}
 \label{1}
 S = \int d^2x\, \left(B^{i\mu\nu}F^i_{\mu\nu}+
\frac{1}{2}A_\mu^i A^{i\mu}\right)
\end{equation}
with $F_{\mu\nu}^i=\partial_\mu A_\nu^i-\partial_\nu A^i_\mu+
f^{ijk}A^j_\mu A^k_\nu$, $i,j,k=1,2,3$, in two dimensional space-time:
$\mu=0,1$. In this case, any antisymmetric tensor $B^{i\mu\nu}$ is
proportional to $\epsilon^{\mu\nu}$, so that
$$B^{i\mu\nu}=\frac{1}{2}B^i\epsilon^{\mu\nu},\quad \epsilon^{01}=1,$$
and the action takes the form
$$ S = \int dx\, \left(\frac{1}{2}B^i\epsilon^{\mu\nu}F^i_{\mu\nu}+
\frac{1}{2}A_\mu^i A^{i\mu}\right).$$
The path integral for this theory reads
\begin{equation}\label{2}
Z = \int DB^i DA_\mu^i\,\exp\{iS\}.
\end{equation}
Let us write down another pair of expressions, formally
equivalent to the above. The first one is derived as follows. Perform
the formal change of variables $A_\mu^i\to(\varphi^i,A^i)$ in~\ref{2},
where
\begin{eqnarray*}
&& A_1^i=\frac{1}{f}f^{ijk}C^{-1}_{kn}(\varphi)\partial_1
C_{nj}(\varphi)\equiv\Lambda_{ji}(\varphi)\partial_1\varphi^j,\\
&& A_0^i=\frac{1}{f}f^{ijk}C^{-1}_{kn}(\varphi)\partial_0
C_{nj}(\varphi)+A^i\equiv\Lambda_{ji}(\varphi)\partial_0\varphi^j+A^i,\\
&& C_{ij}(\varphi)=\left(\exp \hat{\varphi} \right)_{ij},\quad
\hat{\varphi}_{ij}=f^{ijk}\varphi^k,\quad f^{ijk}f^{nji}=f\delta^{in},\\
&& \Lambda_{ij}(\varphi)=\int_0^1 d\tau\,C_{ij}(\tau\varphi)=
\left(\frac{\exp\hat{\varphi}-1}{\hat{\varphi}}\right)_{ij}.
\end{eqnarray*}
When $A^i=0$, the fields $A_\mu^i$ represent a pure gauge. The
Jacobian of the change is
$$\frac{D(A_0^i,A_1^i)}{D(\varphi^j,A^j)} = \prod_x \sqrt{g(\varphi)}\,
\Det\partial_1 \equiv J(\varphi) \Det\partial_1,$$
where $g(\varphi)=\det g_{ij}(\varphi)$,
$g_{ij}(\varphi)=\Lambda_{ik}(\varphi)\Lambda_{jk}(\varphi)$. The path
integral~\ref{2} becomes
\begin{eqnarray*}
&&\kern-16pt Z=\int DB^i D\varphi^i DA^i J(\varphi) \Det\partial_1
\exp\left\{i\int dx\,\left(B^i\nabla_1^{ij}A^j+
\frac{1}{2}\partial_\mu\varphi^i
g_{ij}\partial_\mu\varphi^j+O(A)\right)\right\} \\
&&{}= \int D\varphi^i DA^i J(\varphi)
\Det\partial_1\delta(\nabla_1 A)
\exp\left\{i\int dx\,\left(\frac{1}{2}\partial_\mu\varphi^i
g_{ij}\partial_\mu\varphi^j+O(A)\right)\right\} \\
&& \quad\quad{}= \int D\varphi^i J(\varphi)
{\Det\partial_1 \over \Det\nabla_1}
\exp\left\{i\int dx\,\frac{1}{2}\partial_\mu\varphi^i
g_{ij}\partial_\mu\varphi^j\right\}.
\end{eqnarray*}
Here $\nabla_1^{ij}$ is the covariant derivative
$$\nabla_1^{ij}=\partial_1\delta^{ij}+f^{ikj}A_1^k.$$

One easily checks that
$$\Det\nabla_1^{ij} = \Det\partial_1,$$
which finally leads to
\begin{equation}\label{3}
Z = \int D\varphi^i J(\varphi)
\exp\left\{i\int dx\,\frac{1}{2}\partial_\mu\varphi^i
g_{ij}\partial_\mu\varphi^j\right\}\equiv
\int D\varphi^i J(\varphi) \exp\left\{iS_{\rm ch}(\varphi)\right\}.
\end{equation}
The action $S_{\rm ch}(\varphi)$ is nothing but an action of the model of
principal chiral fields. Hence one concludes that this model is equivalent
to the model described by the Freedman--Townsend action~\ref{1}.

The other expression for $Z$ results after a trivial integration over
the fields $A_\mu^i$:
\begin{eqnarray}
 \label{4}
&& Z   = \int DB^i e^{iS(B)}J_1(B)^{-1/2}, \\
 \label{5}
&& S_{\rm b}(B) =
-\frac{1}{2}\int dx\, \epsilon^{\mu\lambda}\partial_\lambda B^i
N^{-1}_{i\mu|j\nu}\epsilon^{\nu\sigma}\partial_\sigma B^j, \\
&& N^{i\mu|j\nu}=\delta_{ij}\eta^{\mu\nu}+\epsilon^{\mu\nu}
f^{ijk}B^k,  \nonumber \\
&& J_1(B)=\Det N,\quad \eta^{\mu\nu}={\rm diag\,}(1,-1).  \nonumber
\end{eqnarray}
The representation \ref{4} should be interpreted as the partition function
for the theory of scalar fields with the action $S_{\rm b}(B)$~\ref{5}.
So, at the
formal level, one is tempted to accept the equivalence of the theory of
principal scalar fields to the theory~\ref{5}. In the following sections
we show that such a conclusion is incorrect.

\section{Born Amplitudes}\label{2d-Born}
In this section we calculate $2\to2$ scattering amplitudes in the
theory of principal chiral fields, and in the theory described by the
action~\ref{5}.

The action for the principal chiral fields in relevant approximation is
written as
$$ S_{\rm ch}(\varphi)=
\int d^2x\,\left(\frac{1}{2}\partial_\mu\varphi^i
\partial_\mu\varphi^i-\frac{1}{24}f^{ijk}\varphi^j\partial_\mu\varphi^k\cdot
f^{imn}\varphi^m\partial_\mu\varphi^n+O(\varphi^6)\right).$$

The Born $2\to2$ scattering amplitude $A_{\rm ch}$ is given by a
 single diagram
\begin{equation}\label{6}
\vcenter to.76in{\special{em: graph 4.pcx}\vss
\hbox to 1in{\llap{$p_1,i$}\hss$p_3,k$}\vss
\hbox to 1in{\llap{$p_2,j$}\hss$p_4,l$}\vss}
\end{equation}
and equals to
\begin{eqnarray}
 \label{7}
&& A_{\rm ch} = \frac{i}{6}\bigl[f^{ijn}f^{kln}(p_1p_3-p_1p_4)+
         f^{ikn}f^{jln}(p_1p_2-p_1p_4)  \nonumber \\
&&\quad\quad\quad\quad\quad\quad
{}+f^{iln}f^{jkn}(p_1p_2-p_1p_3)\bigr].
\end{eqnarray}
Of course, this expression is valid for any space-time dimension.

The action for $B^i$ field up to an order required is
\begin{eqnarray*}
&& S_{\rm b}(B)=\int d^2x\,\bigl(\frac{1}{2}\partial_\mu B^i
\partial_\mu B^i-\frac{1}{2}\epsilon_{\mu\nu}f^{ijk}B^i\partial_\mu
B^j \partial_\nu B^k\\
&&\quad\quad\quad\quad\quad\quad
{} -\frac{1}{2}f^{ijk}B^j\partial_\mu B^k
f^{iln}B^l\partial_\mu B^n+O(B^5)\bigr).
\end{eqnarray*}
In this theory, the total scattering amplitude $A_B$ equals to a sum
of the amplitudes
$$A_B = A_4+A_{33},$$
where $A_4$ is represented by the diagram \ref{6}, while $A_{33}$ is
given by the sum of diagrams
\begin{equation}\label{8}
\vcenter to.76in{\special{em: graph 33.pcx}\vss
\hbox to 1.5in{\llap{$p_1,a$}\hss$p_3,c$}\vss
\hbox to 1.5in{\llap{$p_2,b$}\hss$p_4,d$}\vss}\qquad\quad
\vcenter to.76in{\special{em: graph 33.pcx}\vss
\hbox to 1.5in{\llap{$p_1,a$}\hss$p_2,b$}\vss
\hbox to 1.5in{\llap{$p_3,c$}\hss$p_4,d$}\vss}\qquad\quad
\vcenter to.76in{\special{em: graph 33.pcx}\vss
\hbox to 1.5in{\llap{$p_1,a$}\hss$p_2,b$}\vss
\hbox to 1.5in{\llap{$p_4,d$}\hss$p_3,c$}\vss}
\end{equation}
The calculations result in
\begin{eqnarray*}
&& A_4 = 12 A_{\rm ch}\\
&& A_{33} = -27 A_{\rm ch},
\end{eqnarray*}
so that
$$ A_B = -15 A_{\rm ch},$$
and thus the Born scattering amplitudes differ in these models. Of course,
any arguments based on calculations of scattering amplitudes may seem
unreliable since the scattering particles are massless and, strictly
speaking, do not exist in two dimensions. More serious arguments will
be given in the section below.

\section{Two-Loop Counterterms}\label{2d-2-loops}

In this section we study the structure of counterterms in the model~\ref{5}.
Both the model of principal chiral fields and the $B^i$-field model are
essentially non-linear. Being renormalizable by their divergence indices
(the counterterm dimensions do not exceed~2), they generally admit an
infinite number of counterterm structures. The existence of a global
symmetry group in the first model allows to prove~\cite{P} that the
renormalization is reduced to a renormalization of the coupling constant
(an overall factor before the total action), and a non-linear renormalization
(reparametrization) of the fields\footnote{Under a suitable choice of
parametrization of the group manifold the renormalization of the fields
becomes multiplicative~\protect\cite{BS,TB}.}. The $B^i$-field model do not
possess any geometric background, thus, no symmetry restrictions on the
choice of imaginable counterterms do exist. However, having taken on truth
the equivalence of the two models, one should expect a renormalization
to be reduced (modulo renormalization of the field $B^i$) solely to the
renormalization of the charge, i.e. the factor before the total action.
In other words, the renormalized action should have the form
\begin{eqnarray}
 \label{9}
&&S_{\rm b}(B)+\frac{\eta}{\epsilon}S_1(B)+\eta^2\left(\frac{1}{\epsilon}
S_{21}(B)+\frac{1}{\epsilon^2}S_{22}(B)\right) \nonumber \\
&&\quad\quad\quad{} =\lambda(\eta,\epsilon)
S_{\rm b}(B)\left({\cal B}(B,\eta,\epsilon)\right)
+O(\eta^3),
\end{eqnarray}
where $\eta$ is the loop expansion parameter (two loops will suit for
what follows),
\begin{eqnarray*}
\lambda(\eta,\epsilon)=1+\frac{\eta}{\epsilon}\lambda_1+
\eta^2\left(\frac{1}{\epsilon} \lambda_{21}+\frac{1}{\epsilon^2}\lambda_{22}
\right),\\
{\cal B}^a(B,\eta,\epsilon)=B^a+\frac{\eta}{\epsilon}F_1^a(B)+
\eta^2\left(\frac{1}{\epsilon}F^a_{21}(B)+\frac{1}{\epsilon^2}F^a_{22}(B)
\right),
\end{eqnarray*}
and $\epsilon$ is the parameter of dimensional regularization we use below.
Comparing the coefficients before $1/\epsilon$ in left and right-hand
sides of the relation~\ref{9}, we see that the following equalities must
hold
\begin{eqnarray}
 \label{10}
&&S_1(B) = \lambda_1 S_{\rm b}(B)+\frac{\delta S_{\rm b}(B)}{\delta
B^i}F_1^i(B),\\
 \label{11}
&&S_{21}(B) = \lambda_{21} S_{\rm b}(B)+
\frac{\delta S_{\rm b}(B)}{\delta B^i}F_{21}^i(B).
\end{eqnarray}
We restrict ourselves to the case of {\sl SU(2)\/} group, when the action
$S_{\rm b}(B)$ can be written out explicitly. After the transition to Euclidean
space-time, it reads
\begin{equation}\label{12}
S_{\rm b}(B)=\frac{1}{2}\int d^2x\,\partial_\mu B^i\left[
\frac{\delta_{ij}+B^iB^j}{1+B^2}\delta_{\mu\nu}+ i\epsilon^{\mu\nu}
\frac{\epsilon^{ijk}B^k}{1+B^2}\right]\partial_\nu B^j.
\end{equation}
Due to conservation of global {\sl SU(2)\/} group, the renormalizations have
the following general structure
\begin{eqnarray*}
&& S_1(B)=\frac{1}{2}
\int d^2x\,\partial_\mu B^i\bigl[\delta_{\mu\nu}\delta_{ij}A_1(z)+
\delta_{\mu\nu}B^iB^jC_1(z)\\
&&\quad\quad\quad\quad{}+i\epsilon^{\mu\nu}\epsilon^{ijk}B^kD_1(z)\bigr]
\partial_\nu B^j,\\
&& S_{21}(B)=\frac{1}{2}
\int d^2x\,\partial_\mu B^i\bigl[\delta_{\mu\nu}
\delta_{ij}A_{21}(z)+ \delta_{\mu\nu}B^iB^jC_{21}(z)\\
&&\quad\quad\quad\quad{}+i\epsilon^{\mu\nu}\epsilon^{ijk}B^kD_{21}(z)\bigr]
\partial_\nu B^j,\\
&& F_{1}^i(B)=B^if_{1}(z),\quad F_{21}^i(B)=B^if_{21}(z),\quad
   z = B^iB^i\equiv B^2.
\end{eqnarray*}
Eq.~\ref{10},\ref{11} imply the relations between the functions introduced
above:
\begin{eqnarray}
 \label{13}
&&
A_{1}(z) = \frac{2f_{1}(z)}{(1+z)^2}+\frac{\lambda_{1}}{1+z},\quad
D_{1}(z) = \frac{(3+z)f_{1}(z)}{(1+z)^2}+\frac{\lambda_{1}}{1+z},
\nonumber \\
&&
C_{1}(z) = \frac{4+2z}{(1+z)^2}f_{1}(z)+4f_{1}'(z)+
\frac{\lambda_{1}}{1+z},\\
 \label{14}
&&
A_{21}(z) = \frac{2f_{21}(z)}{(1+z)^2}+\frac{\lambda_{21}}{1+z},\quad
D_{21}(z) = \frac{(3+z)f_{21}(z)}{(1+z)^2}+\frac{\lambda_{21}}{1+z},
\nonumber \\
&&
C_{21}(z) = \frac{4+2z}{(1+z)^2}f_{21}(z)+4f_{21}'(z)+
\frac{\lambda_{21}}{1+z}.
\end{eqnarray}

Below we present the results of calculations of the functions
$A_1$, $C_1$, $D_1$, $A_{21}$, $C_{21}$, $D_{21}$, and the solutions to
the equations \ref{13},\ref{14}.

It is known~\cite{F1,TB2} that up to a non-linear change of variables,
the counterterms are expressed in terms of the geometric objects, namely,
through the metric, curvature, torsion and their covariant derivatives.
In our case the action may be rewritten in the form (in Euclidean
space-time)
$$ S_{\rm b}(B)=\frac{1}{2}\int d^2x\,
\partial_\mu B^i\left(g_{ij}(B)\delta_{\mu\nu}+i\epsilon^{\mu\nu}
h_{ij}(B)\right)\partial_\nu B^j,$$
where
$$g_{ij}(B)=\frac{\delta_{ij}+B_iB_j}{1+B^2},\quad
 h_{ij}=\frac{\epsilon^{ijk}B^k}{1+B^2}.$$

We will need the following expressions:

\begin{eqnarray}\label{15a}
&& g^{ij}= ( 1 + B^2 )\delta_{ij}  - B^i B^j,\nonumber \\
&& \Gamma^i_{jk} = \frac{1}{2} g^{in}
(\partial_j g_{nk}+\partial_k g_{jn}-\partial_n g_{jk} )
\nonumber \\
&&\quad{}=  {  ( 2 + B^2 ) B^i \delta_{jk} -  ( 1 + B^2 ) B^j \delta_{ik} -
       ( 1 + B^2 ) B^k \delta_{ij} +  B^i B^j B^k \over (1+B^2)^2 },
\nonumber  \\
&& H_{ijk} =
\partial_k h_{ij}+\partial_j h_{ki}+\partial_i h_{jk} \nonumber \\
&& \qquad\qquad{}={ 3 + B^2 \over (1+B^2)^2 } \epsilon_{ijk},\nonumber  \\
&& \Ten{R}^i_{jkl} =
\partial_k \Gamma^i_{jl}-\partial_l \Gamma^i_{jk}+
\Gamma^i_{nk}\Gamma^n_{jl}-\Gamma^i_{nl}\Gamma^n_{jk} \\
&&\quad{}=
(-3 \delta_{il} \delta_{jk} - 3 B^2 \delta_{il} \delta_{jk} - B^4 \delta_{il}
\delta_{jk} +   3 \delta_{ik} \delta_{jl} +3 B^2 \delta_{ik} \delta_{jl} +
B^4 \delta_{ik} \delta_{jl} \nonumber \\
&&\quad\quad\quad\quad\quad\quad{}-
 3 \delta_{jl} B^i B^k - B^2 \delta_{jl} B^i B^k
 +B^2 \delta_{il} B^j B^k + 3 \delta_{jk} B^i B^l
\nonumber  \\ &&\quad\quad\quad\quad\quad{}
 +B^2 \delta_{jk} B^i B^l - B^2 \delta_{ik} B^j B^l)/ (1 + B^2)^3.\nonumber
\end{eqnarray}

\subsection{One-Loop Approximation}
The one-loop counterterm (modulo the change of fields) equals~\cite{Ctz}
\begin{eqnarray*}
&& S_1(B) = -\frac{1}{2}\int d^2x\, \partial_\mu B^i\left(
\hat{R}_{(ij)}+i\epsilon^{\mu\nu}\hat{R}_{[ij]}\right)\partial_\nu B^j\\
&& \hat{R}_{(ij)}=\frac{1}{2\pi}
\left(R_{ij}-\frac{1}{4}H_{imn}H_j^{mn}\right),\\
&& \hat{R}_{[ij]}=-\frac{1}{4\pi} D^k H_{ijk}.
\end{eqnarray*}
The calculation gives
\begin{eqnarray*}
&& \hat{R}_{(ij)}= - \frac{3+B^4}{4\pi(1+B^2)} \delta_{ij} +
\frac{3+8B^2+B^4}{4\pi(1+B^2)^3}B^iB^j,\\
&& \hat{R}_{[ij]}= - \frac{1}{\pi} \frac{1}{(1+B^2)}\epsilon^{ijk}B^k.
\end{eqnarray*}
The equations~\ref{13} in this case have the solution
$$ f_1(z) = -\frac{1-B^2}{4\pi(1+B^2)}, \quad \lambda_1=-\frac{1}{4\pi}.$$
So, the renormalization of the model with the action \ref{12} is reduced
to the renormalization of a single parameter (the factor before the
action) exactly as it has been in a chiral theory; the renormalizations
in these models are, however, different (chiral theory had
$\lambda_1=-1/\pi$ \cite{P0,B}).

\subsection{Two-Loop Approximation}
The metric renormalization will turn out to be sufficient for our
purposes, i.e.\ only the function ${\beta_{g}^{(2)}}{}_{ij}(B)$ in the
expression for the two loop counterterm
$$ S_{21}(B)=\frac{1}{2} \int d^2x\, \partial_\mu B^i
\left(-\frac{1}{2}{\beta^{(2)}_g}{}_{ij}(B)\delta_{\mu\nu}
-\frac{i}{2}{\beta^{(2)}_h}{}_{ij}(B)
\epsilon_{\mu\nu}\right)\partial_\nu B^j$$
matters.

The expression for ${\beta^{(2)}_g}{}_{ij}$ we used was taken from the
work~\cite{Mz} (see also~\cite{Z})
\begin{eqnarray} \label{16}
&&{\beta^{(2)}_g}{}_{ij}   = \frac{1}{4\pi^2} \biggl\{\frac{1}{2}\biggl[
R_{iabc}\Ten{R}_j\u^{abc}      
-\frac{3}{2}\Ten{R}_{(i}\u^{abc}H_{j)al} 
\Ten{H}_{bc}\u^l                                       
-\frac{1}{2}R^{abrs}H_{iab}H_{jrs} 
+\frac{1}{8}(H^4)_{ij}                                     
\nonumber\\
&&\quad{}
+\frac{1}{4}{\cal D}_l H_{iab} {\cal D}^l 
\Ten{H}_j\u^{ab}                                      
+\frac{1}{12}{\cal D}_i H_{abc}                  
{\cal D}_j H^{abc}                               
+\frac{1}{8}H_{ial}                                 
\Ten{H}_{jb}\u^l(H^2)^{ab}                  
\biggr]
\nonumber\\
&&\quad{}+p_1\biggl[
R_{iabj}(H^2)^{ab}                         
+2\Ten{R}_{(i}\u^{abc}H_{j)al}
\Ten{H}_{bc}\u^{l}+
R^{abrs}H_{abi}H_{rsj}
\nonumber\\
&&\quad\quad{}
- {\cal D}_l H_{iab}{\cal D}^l
\Ten{H}_j\u^{ab}\biggr]+p_1{\cal D}_{(i}
({\cal D}^l H^2_{j)l}-\frac{1}{2}{\cal D}_{j)}H^2)
\nonumber\\
&&{} + p_2 H_{ab(i}
\left({\cal D}_{j)}{\cal D}_l H^{lab}+
2{\cal D}^a{\cal D}_l \Ten{H}_{j)}\u^{lb}\right)
\biggr\}.
\end{eqnarray}
Here
\begin{eqnarray*}
&& (H^2)_{ij} \equiv H_{iab}\Ten{H}_j\u^{ab}, \\
&& (H^4)_{ij} \equiv \frac{1}{2}H_{ial}H^{arb}
\Ten{H}_{rs}\u^l H^s\u_{jb}
+ (i\leftrightarrow j).
\end{eqnarray*}
The coefficients $p_1$ and $p_2$ are arbitrary; they reflect the freedom
in the definition of $\epsilon^{\mu\nu}$ in dimensional regularization, and
the possibility of finite metric renormalizations~\cite{Mz}.

The calculation of separate terms of the expression~\ref{16} gives
\begin{eqnarray*}
&&t_1\equiv R_{iabc} R_j^{abc} =
2 (18 \delta_{ij} + 18 B^2 \delta_{ij} + 15 B^4 \delta_{ij} +
6 B^6 \delta_{ij} +
 B^8 \delta_{ij} \\
&&{}\qquad\qquad{} -15 B^2 B^i B^j - 6 B^4 B^i B^j - B^6 B^i B^j)/
  (1 + B^2)^5 \\
&&t_2\equiv (-3/2) R_i^{abc} H_{jal} H_{bc}^l =
3 (-54 \delta_{ij} - 63 B^2 \delta_{ij} - 33 B^4 \delta_{ij} -
9 B^6 \delta_{ij} -
      B^8 \delta_{ij} \\
&&{}\qquad\qquad{} -27 B^i B^j - 9 B^2 B^i B^j + 3 B^4 B^i B^j +
      B^6 B^i B^j)/(1 + B^2)^5 \\
&&t_3\equiv (-1/2) R^{abrs} H_{iab} H_{jrs} =
-2 (27 \delta_{ij} + 18 B^2 \delta_{ij} + 3 B^4 \delta_{ij}\\&&{}\qquad
 + 54 B^i B^j +
      72 B^2 B^i B^j + 39 B^4 B^i B^j + 10 B^6 B^i B^j +
      B^8 B^i B^j)/(1 + B^2)^5 \\
&&t_4\equiv  (1/8) (H^4)_{ij} =
(81 \delta_{ij} + 108 B^2 \delta_{ij} + 54 B^4 \delta_{ij} + 12 B^6
\delta_{ij} +  B^8 \delta_{ij} \\
&&{}\qquad + 81 B^i B^j + 108 B^2 B^i B^j + 54 B^4 B^i B^j +
    12 B^6 B^i B^j + B^8 B^i B^j)/4 (1 + B^2)^5 \\
&& t_5\equiv (1/4) D_{l}H_{iab} D^{l}H_j^{ab} =
8 B^2 (\delta_{ij} + B^i B^j)/(1 + B^2)^5  \\
&& t_6\equiv (1/12) D_{i}H_{abc} D_{j}H^{abc} =
 8 B^i B^j/(1 + B^2)^4 \\
&& t_7\equiv  (1/8) H_{ial} H_{jb}^{l} (H^2)^{ab} =
 (81 \delta_{ij} + 108 B^2 \delta_{ij} + 54 B^4 \delta_{ij} +
12 B^6 \delta_{ij} +
    B^8 \delta_{ij}\\
&&{}\qquad + 81 B^i B^j + 108 B^2 B^i B^j + 54 B^4 B^i B^j +
    12 B^6 B^i B^j + B^8 B^i B^j)/2 (1 + B^2)^5 \\
&& t_8\equiv R_{iabj} (H^2)^{ab} =
2 (-54 \delta_{ij} - 63 B^2 \delta_{ij} - 33 B^4 \delta_{ij} -
9 B^6 \delta_{ij} -
B^8 \delta_{ij}\\
&&{}\qquad\qquad - 27 B^i B^j - 9 B^2 B^i B^j + 3 B^4 B^i B^j +
B^6 B^i B^j)/(1 + B^2)^5 \\
&& t_9\equiv H_{kli} ( D_j(D_m(H^{mkl})) + 2 D^k(D_m(H^{ml}_j))) =
 24 (-3 \delta_{ij} + 2 B^2 \delta_{ij} + B^4 \delta_{ij} \\
&&{}\qquad\qquad - 3 B^i B^j +
       2 B^2 B^i B^j + B^4 B^i B^j)/(1 + B^2)^5
\end{eqnarray*}

Thus, the final expression for ${\beta^{(2)}_g}{}_{ij}$ is
\begin{eqnarray}\label{18}
&& {\beta^{(2)}_g}{}_{ij}
 = \frac{1}{4\pi^2}\biggl(\frac{1}{2}(t_1+t_2+t_3
 + t_4+t_5+t_6+t_7) \nonumber \\
&& \quad\quad{}+ p_1\left(t_8-\frac{4}{3}t_2-2 t_3
     -4 t_5\right)+p_2 t_9\biggr)\nonumber \\
&&{} = ( (-477 + 1728p_1 - 576p_2 - 400B^2 +
    1328p_1B^2 + 384p_2B^2 - 138B^4\nonumber \\
&&{}  + 624p_1B^4 + 192p_2B^4 - 24B^6 +
    144p_1B^6 - B^8 + 16p_1B^8 ) \delta_{ij}\nonumber \\
&&{}  + ( -481 + 2160p_1 - 576p_2 - 416B^2 +
    2192p_1B^2 + 384p_2B^2 - 162B^4 \nonumber\\
&&\qquad{}  + 1200p_1B^4 + 192p_2B^4 - 40B^6 +
    304p_1B^6 - 5B^8 \nonumber\\
&&\qquad\qquad{} + 32p_1B^8) B_iB_j )/(32\pi^2(1 + B^2)^5)
\end{eqnarray}
As noted in~\cite{Mz}, for $p_1=1/4$, $p_2=0$, the expression~\ref{16}
coincides with the symmetric part of
\begin{eqnarray} \label{19}
&&
\hat{\beta}^{(2)}_g{}_{ij} =
\frac{1}{8\pi^2} \biggl(\hat{R}^{abc}\!\vphantom{R}_{(j}
\hat{R}_{i)abc} - \frac{1}{2}
\hat{R}^{bca}\!\vphantom{R}_{(j}
\hat{R}_{i)abc}+\frac{1}{2}\hat{R}_{a(ij)b}
(H^2)^{ab}\biggr)
\nonumber\\
&&\qquad\qquad \equiv
\frac{1}{4\pi^2} (\hat{t}_1+\hat{t}_2+\hat{t}_3),
\end{eqnarray}
where $\hat{R}^a_{bcd}$ is given by its standard formula twisted
with the change
$$ \hat{\Ten{\Gamma}}^l\!\u_{ij} = \Gamma^l\!\u_{ij}-
   \frac{1}{2}\Ten{H}^l\!\u_{ij}.$$
To cross-check the validity of calculations, we have also evaluated the
expression~\ref{19}:
\begin{eqnarray*}
&&\hat{t}_1 = ( ( 9 - 32 B^2 + 42 B^4 + 24 B^6 + 5 B^8) \delta_{ij} +
       (24 - 48 B^2 -  8 B^4) \epsilon_{ijk}B^k \\
&&\quad\quad{}+ (69 + 16 B^2 + 18 B^4 +  8 B^6 + B^8) B^i B^j )/8
  (1 + B^2)^5, \\
&&\hat{t}_2 = ( ( 9 - 32 B^2 + 42 B^4 + 24 B^6 + 5 B^8) \delta_{ij} +
       (24 - 48 B^2 -  8 B^4) \epsilon_{ijk}B^k   \\
&&\quad\quad{}+        (-59 -112 B^2 + 18 B^4 +  8 B^6 + B^8) B^i B^j )/
8 (1 + B^2)^5, \\
&&\hat{t}_3 = (-(27 + 18 B^2 + 12 B^4  + 6 B^6 +   B^8) \delta_{ij} -
       (36 + 24 B^2 +  4 B^4) \epsilon_{ijk}B^k     \\
&&\quad\quad{}+ (27 + 90 B^2 + 60 B^4 + 14 B^6 + B^8) B^i B^j )/
2 (1 + B^2)^5, \\
&&
\hat{\beta}^{(2)}_g{}_{ij} =
 ((-45 - 68 B^2 + 18 B^4 + 12 B^6 + 3 B^8) \delta_{ij} \\
&&\qquad\qquad{}- (48 + 96 B^2 + 16 B^4) \epsilon_{ijk}B^k     \\
&&\quad\quad{}+ (59 +132 B^2 +138 B^4 + 36 B^6 + 3 B^8)
B^a B^b )/32\pi^2(1 + B^2)^5.
\end{eqnarray*}
Its symmetric part is seen to coincide with~\ref{18} for $p_1=1/4$, $p_2=0$.

The functions $A_{21}(z)$ and $C_{21}(z)$ turn out to be
\begin{eqnarray*}
A_{21}(z) &=&
 (477 - 1728 p_1 + 576 p_2 + 400 z - 1328 p_1 z - 384 p_2 z \\
&& {} + 138 z^2 -
  624 p_1 z^2 - 192 p_2 z^2 + 24 z^3 - 144 p_1 z^3\\
&& {}  + z^4 - 16 p_1 z^4)/ 64\pi^2 (1 + z)^5, \\
C_{21}(z) &=&
 (481 - 2160 p_1 + 576 p_2 + 416 z - 2192 p_1 z - 384 p_2 z \\
&&\quad{} + 162 z^2 -  1200 p_1 z^2 - 192 p_2 z^2 + 40 z^3 - 304 p_1 z^3  \\
&&\quad\quad+ 5 z^4 - 32 p_1 z^4)/ 64\pi^2 (1 + z)^5.
\end{eqnarray*}
{}From the Eq.~\ref{14} which includes the function $A_{21}(z)$, one finds
\begin{eqnarray*}
f_{21}(z)& = &
 (477 + 8 \lambda_{21} - 1728 p_1 + 576 p_2 + 400 z +
 32 \lambda_{21} z - 1328 p_1 z -
 384 p_2 z\\
&& \quad\quad{} + 138 z^2 + 48 \lambda_{21} z^2 - 624 p_1 z^2 -
 192 p_2 z^2 \\
&& \qquad\qquad\qquad{} + 24 z^3 + 32 \lambda_{21} z^3 - 144 p_1 z^3 \\
&&\qquad {}+ z^4 + 8 \lambda_{21} z^4 -  16 p_1 z^4)/128\pi^2 (1 + z)^3.
\end{eqnarray*}
Substituting this into the Eq.~\ref{14}, where $C_{21}(z)$ enters, and
taking the derivative $f_{21}'(z)$ under assumption $\lambda_{21}={\rm
const}$, we get
\begin{eqnarray*}
&&\lambda_{21}(z) =
(-1589 + 6416 p_1 - 3648 p_2 - 660 z + 1920 p_1 z  \\
&&\qquad{} +384 p_2 z - 6 z^2 -  96 p_1 z^2 + 192 p_2 z^2 \\
&&{}  + 28 z^3 - 256 p_1 z^3 + 3 z^4 - 48 p_1 z^4)/
  (192 \pi^2 (1 + z)^4) \\
&&{} =  {-1589 + 6416 p_1 - 3648 p_2 \over  192\pi^2 } +
  {89  - 371 p_1 + 234 p_2 \over 3 \pi^2    } z \\
&&\quad{}  +
  {-1657 + 7048 p_1 - 4728 p_2 \over 24 \pi^2 } z^2 +
  {1577 - 6812  p_1 + 4752 p_2 \over 12\pi^2 } z^3 + O(z^4).
\end{eqnarray*}
The condition of vanishing of the coefficients before $z$ and $z^2$
gives:
$$p_1 = \frac{5509}{17476},\quad p_2 = \frac{4175}{34952}.$$
Under that, the coefficient before $z^3$ is non-zero and equals to
$997/8738\pi^2$.

Thus, no choice of $p_1$ and $p_2$ might make $\lambda_{21}$ constant.
It means that the $B$-field action~\ref{12} corresponds to a theory
with multiple (finite or infinite) number of coupling constants, and hence
{\sl can not\/} be equivalent to the model of principal chiral fields where
this number is one.

\section{3-Dimensional Model}\label{3d-Born}
In this section we calculate $2\to2$ Born scattering amplitude for
the three dimensional Freedman--Townsend model. It is described by the
action
$$ S = \frac{1}{2} \int \left( B_\mu^i \epsilon^{\mu\nu\lambda}
F^i_{\nu\lambda}+A_\mu^i A^{i\mu}\right)d^3x,$$
where $\epsilon^{\mu\nu\lambda}$ is the totally antisymmetric tensor,
$\epsilon^{012}=1$.

This is a gauge theory. The gauge transformations read
$$\delta A_\mu^i=0,\quad
\delta B_\mu^i=(\partial_\mu\delta_j^i+f^{ikj}A^k_\mu)\xi^j.$$

Formally, the model is again equivalent to the model of principal chiral
fields. The simplest way to convince oneself in it---is to evaluate
the path integral in a definite gauge. Choosing for example $B_2^i=0$,
which requires the Faddeev-Popov determinant
$\Delta=\Det|\partial_2\delta_j^i+f^{ijk}A^k|$, the integral over
$B_\mu^i$ gives
$$\delta(G_{12})\delta(G_{02}).$$
After the change of integration variables $A_\mu^i\to\varphi^i,a_0^i,a_1^i$
$$A_2^i=\Lambda_{ji}\partial_2\varphi^j,\quad
  A_0^i=\Lambda_{ji}\partial_0\varphi^j+a_0^i,\quad
  A_1^i=\Lambda_{ji}\partial_1\varphi^j+a_1^i,$$
we get, in analogy with section~\sref{Formal}, the same expression~\ref{3}
for $Z$. Correspondingly, the Born scattering amplitude of two particles
$\varphi$ is given by Eq.~\ref{7}.

On the contrary, integrating over $A_\mu^i$ first will produce the theory
of $B_\mu^i$ fields with the action
\begin{eqnarray*}
&& S_{\rm b}(B) = -\frac{1}{2}\int d^3x\,\epsilon^{\mu\lambda\sigma}
\partial_\lambda B_\sigma^i N^{-1}_{i\mu|j\nu}\epsilon^{\nu\gamma\delta}
\partial_\gamma B_\delta^j \\
&&{}=\frac{1}{2}\int d^3x\,\bigl[B_\mu^i(\Box\eta_{\mu\nu}-
\partial_\mu\partial_\nu)B_\nu^i-\frac{1}{2}\epsilon^{\mu\nu\lambda}
f^{abc}B^i_{\nu\lambda}B^j_\sigma B^k_{\sigma\mu} \\
&&\qquad\qquad
{}- f^{ijk} B_{\mu\nu}^j B_\nu^k f^{iln} B_{\mu\lambda}^l B_\lambda^n
\bigr]+O(B^5),\\
&&N_{i\mu|j\nu}=\eta_{\mu\nu}\delta^{ij}+\epsilon^{\mu\nu\lambda}
f^{ijk}B_\lambda^k,\quad B^i_{\mu\nu}\equiv \partial_\mu B_\nu^i-
\partial_\nu B_\mu^i.
\end{eqnarray*}
Of course, this theory is still a gauge theory; to gauge-fix it we add
$$-\frac{1}{2}\int d^3x\,\partial_\mu B_\mu^i\,\partial_\nu B_\nu^i$$
to the action. The ghosts action doesn't matter for what follows. The
scattering amplitude is expressed through the vertex function
$\Gamma_{\mu\nu\lambda\sigma}^{ijkl}$ as
$$A=\xi_\mu(p_1)\xi_\nu(p_2)\xi_\lambda(p_3)\xi_\sigma(p_4)
\Gamma_{\mu\nu\lambda\sigma}^{ijkl}(p_1,p_2,p_3,p_4),$$
where all $p_n$ are taken on mass shell, $p_n^2=0$, $\sum p_n=0$, and
$\xi_\mu(p)$ is the polarization vector for a physical state
$$\xi_\mu=\frac{1}{p_0}\epsilon^{0\mu\nu}p_\nu.$$
The diagram of the type~\ref{6} gives the following contribution to the
scattering amplitude
\begin{eqnarray*}
&& A_4=\frac{i}{p_{10}p_{20}p_{30}p_{40}}\biggl\{
f^{ijn}f^{kln}\bigl[(p_1p_2)^2(p_{10}p_{40}+p_{20}p_{30}-p_{10}p_{30}-
p_{20}p_{40}) \\
&& \qquad{}+(p_1p_3)(p_{10}^2p_{20}p_{40}+p_{10}p_{30}p_{40}^2+
p_{20}p_{30}^2p_{40}
+p_{10}p_{20}^2p_{30}-p_{10}^2p_{30}^2-
p_{20}^2p_{40}^2\\
&& \qquad\qquad{}-p_{10}^2p_{20}p_{30}
-p_{20}p_{30}p_{40}^2-p_{10}p_{30}^2p_{40}-p_{10}p_{20}^2p_{40}) \\
&& \qquad{}-(p_1p_4)(p_{10}p_{20}^2p_{40}+p_{10}^2p_{20}p_{30}
+p_{20}p_{30}p_{40}^2 +p_{10}p_{30}^2p_{40}
-p_{20}^2p_{30}^2-p_{10}^2p_{40}^2 \\
&& \qquad\qquad{}-p_{10}p_{20}^2p_{30}
-p_{10}^2p_{20}p_{40}-p_{10}p_{30}p_{40}^2-
p_{20}p_{30}^2p_{40})\bigr] \\
&&\qquad\qquad{}
+(j,2\leftrightarrow k,3)
+(j,2\leftrightarrow l,4)\biggr\}.
\end{eqnarray*}
The diagrams of the type~\ref{8} contribute as
\begin{eqnarray*}
&& A_{33} = \frac{i}{p_{10}p_{20}p_{30}p_{40}} \biggl\{f^{ijn}f^{kln}\bigl[
(p_1p_2)^2(p_{10}p_{40}+p_{20}p_{30}-p_{10}p_{30}-p_{20}p_{40}) \\
&&\qquad{}+(p_1p_4)(p_{30}p_{40}-p_{10}p_{20} -
p_{10}^2-p_{20}^2)(p_{10}p_{40}+p_{20}p_{30}) - (p_1p_3)(p_{30}p_{40} \\
&&\qquad{}-p_{10}p_{20}-p_{10}^2-p_{20}^2)(p_{10}p_{30}+
p_{20}p_{40})-\frac{(p_1p_3)-(p_2p_4)}{2}p_{10}p_{20}p_{30}p_{40}\bigr]\\
&&\qquad\qquad
+(j,2\leftrightarrow k,3)
+(j,2\leftrightarrow l,4)\biggr\}.
\end{eqnarray*}
Summing up these expressions, we finally get
\begin{eqnarray*}
&& A=A_4+A_{33}=\frac{i}{6} \bigl[f^{12i}f^{34i}(p_1p_3-p_1p_4)+
f^{13i}f^{42i}(p_1p_4-p_1p_2) \\
&&\qquad\qquad{}+ f^{14i}f^{23i}(p_1p_2-p_1p_3)\bigr],
\end{eqnarray*}
which is nothing but a scattering amplitude in the theory of the principal
chiral fields (see~\ref{7}).

This example confirms the hypothesis that the two dimensional anomaly
discussed above is caused by severe infrared singularities of massless
fields.

{\bf Acknowledgement.}
The authors are grateful to B.L.Voronov, R.R.Metsaev and A.A.Tseytlin for
interesting remarks and discussions.

A.S was supported in part by the Russian Foundation of Fundamental
Investigations, Grant \# 93--02--15541,
I.V.T was supported in part by the International Science Foundation
under Grant {\sl N}~ M2I000 and by the European Community Commission
under the contracts INTAS--93--2058 and INTAS--93--633.

\end{document}